\documentclass[10pt]{article}
\usepackage{graphicx, lipsum, comment, caption, amsmath, amssymb, hyperref, comment}
\usepackage[margin=1in]{geometry}
\usepackage[T1]{fontenc}
\usepackage[utf8]{inputenc}

\title{Molecular-scale, nonlinear actomyosin binding dynamics drive population-scale adaptation and evolutionary convergence}

\author
{Jake McGrath$^{1}$, Colin Johnson$^{1}$, Jos\'{e} Alvarado$^{1 \ast}$\\
\\
\normalsize{$^{1}$Center for Nonlinear Dynamics, Department of Physics,} \\[0.5ex] 
\normalsize{The University of Texas at Austin. Austin, TX, USA}\\\\
\normalsize{$^{\ast}$ : To whom correspondence should be addressed; E-mail: alv@chaos.utexas.edu}
}
\date{\today}

\begin{document}

\maketitle

\subsection{Abstract}

Biological actuators --- from myosin motors to muscles --- follow Hill’s model where a dimensionless parameter $\alpha$ captures the nonlinear coupling between contraction rate and force generation. 
Our prior work identified a characteristic $\alpha^* = 3.85 \pm 2.32$ across natural muscles and showed that $\alpha^*$ optimizes a power-efficiency tradeoff, potentially explaining its prevalence in nature. 
However, those results reflected short-term actuation tasks whereas phenotypic distributions in $\alpha$ emerge over evolutionary timescales. 
Here, we use numerical simulations of self-propelled agents to explore how nonlinear actomyosin actuation (parameterized by $\alpha$) shapes population dynamics. 
Agents of different $\alpha$ compete for resources and reproduce with slight mutations. 
Without mutations, resource availability drives populations in $\alpha$ toward distinct behaviors: under abundance or scarcity, specialized $\alpha$ survive. 
However, with mutations and selection, populations evolve toward distributions centered around the characteristic $\alpha^*$ observed in nature. 
Further, we show that the mutation rate $\delta$ governs a balance between adaptability and robustness: large $\delta$ generates instability and extinction, small $\delta$ prevents feedback, while intermediate $\delta$ enables long-term adaptability while remaining robust to short-term noise. 
Our results suggest that nonlinear actuation provides a general understanding of energy management in actomyosin systems across a wide range of timescales, ranging from the task-specific to evolutionary.
These insights may guide the rational design of active materials with adaptive properties.

\section{Introduction}

Who wins the race, the tortoise or the hare \cite{aesop_aesops_2002}? 
The answer depends not only on the racers' inherent actuation properties, but also on their environment. 
Over short distances, the hare’s speed provides a clear advantage, whereas over longer races the tortoise’s efficiency becomes decisive. 
Speed-efficiency tradeoffs clearly help with predicting the outcome of a single race. 
But if races were repeated over many generations and racers could adjust their actuation strategies, those balancing speed and efficiency might ultimately be favored. 
Additional factors such as mutation rate, resource availability, and competition help determine which strategies are the most fit. 
Yet quantifying how these factors determine speed-efficiency tradeoffs and hence survival is difficult: improving along one performance axis often compromises another.

Tradeoffs --- fitness costs incurred when improvement of one trait degrades another --- are a fundamental consequence of evolution \cite{stearns_trade-offs_1989, garland_trade-offs_2022, mauro_trade-offs_2020}. 
They arise because selection often acts on competing components of fitness, preventing traits from reaching theoretical extremes \cite{stearns_trade-offs_1989}, which are further constrained by physical, physiological, or developmental limits \cite{garland_trade-offs_2022}. 
Because organisms perform myriad tasks and no phenotype can optimize all functions, evolution is therefore constrained to Pareto-optimal sets where gains along one axis incur losses along another \cite{shoval_evolutionary_2012}. 
Classic examples include armored animals, such as tortoises or armadillos, which gain protection from predators but incur slower locomotion and metabolic rates \cite{superina_life_2012, golubovic_locomotor_2017}, and functional tradeoffs in locomotion, such as cheetahs, which maximize sprinting speed through their anatomy at the cost of climbing ability, versus leopards, which trade land-speed for arboreal-agility \cite{noor_efficiency_2012}. 
Similar patterns emerge across scales: in bacteria, adaptation to high temperatures reduces cold tolerance \cite{rodriguez-verdugo_different_2014}, and in yeast, no single mutation can simultaneously optimize fermentation, respiration, and stationary-phase performance \cite{li_single_2019}.
These insights highlight that tradeoffs are a core organizing principle of evolution across diverse organisms and scales.

When disparate populations experience similar constraints and tradeoffs, evolution often produces convergent outcomes, yielding comparable functional solutions. 
Examples range from morphological convergence --- wings in bats and birds \cite{maina_what_2000}, streamlined bodies in aquatic vertebrates \cite{davis_locomotion_2019, fish_transitions_1996}, and camera-type eyes in cephalopods and vertebrates \cite{nilsson_cephalopod_2023, yoshida_molecular_2015} --- often shaped by physical limits \cite{kempes_scales_2019, liu_comprehensive_2025}, to quantitative scaling relationships between body mass and metabolic rate, lifespan, growth, and locomotor energetics \cite{west_general_1997, west_origin_2005, hatton_linking_2019, clemente_how_2023, bastille-rousseau_allometric_2016}. 
Behavioral convergence likewise emerges, for instance in predator move–wait timing \cite{wearmouth_scaling_2014}. 
Such repeated solutions indicate that evolutionary pathways are constrained and partially predictable under shared competitive interactions, physical limits, and developmental limits \cite{mcghee_convergent_2011}, effectively funneling adaptation toward regions of trait space.

Muscle is an example that illustrates quantitative convergence across evolutionarily distant species. 
As a biological actuator, muscle generates force through interactions between actin filaments and myosin motors: myosin heads hydrolyze ATP to produce force while actin provides structure \cite{alberts_molecular_2002}. 
The dynamics of these molecular motors are inherently nonlinear: myosin heads detach from actin at rates that increase with sliding velocity \cite{piazzesi_skeletal_2007}. 
When aggregated across many motors, this microscopic nonlinearity gives rise to the macroscopic nonlinear force–velocity (FV) relationship observed in muscle \cite{hill_heat_1938, piazzesi_skeletal_2007, seow_hills_2013}, in which muscle force decreases nonlinearly with contraction speed. 
Remarkably, muscles across taxa exhibit a conserved concave FV relationship whose curvature --- often expressed as the dimensionless parameter $\alpha$ --- is consistently measured to be $\alpha^* \approx 4$ across muscles in evolutionarily distant species \cite{mcgrath_microscale_2025}. 
Such conservation suggests that global physical constraints and tradeoffs guide muscle evolution across taxa. 
Previous work showed that this curvature can emerge from minimizing a tradeoff between a muscle's energetic efficiency and mechanical power output \cite{mcgrath_hill-type_2022, mcgrath_microscale_2025}. 
However, those studies focused on short-term performance, leaving open the question of how $\alpha$ dynamically evolves under performance tradeoffs over evolutionary timescales and why the specific nonlinearity $\alpha^* \approx 4$ consistently emerges.

To address this question, we employ an agent-based model (ABM) to simulate evolutionary dynamics resulting from competitive interactions among individuals. 
By representing individuals as discrete agents whose reproduction, mutation, and interactions generate population-level outcomes, ABMs capture dynamics often intractable to analytical methods, particularly in the weak-selection, strong-mutation regime \cite{adami_evolutionary_2016}. 
Across biological systems, ABMs have been used to show how competition, mutation, and selection shape genetic structure \cite{lamarins_importance_2022}; and how mating interactions drive co-evolution of traits and genetic architectures over long timescales \cite{chevalier_demogenetic_2022}. 
Critically, ABMs have also reproduced key experimental phenomena, including punctuated adaptation, epistasis, and historical contingency in Lenski’s long-term \textit{E. coli} experiment \cite{root-bernstein_evolutionary_2024, lenski_long-term_1991}. 
At larger ecological scales, ABMs have been used to model multi-species interactions and local adaptation across spatial landscapes \cite{oddou-muratorio_integrating_2025}. 
Collectively, these studies highlight ABMs as a mechanistic framework for exploring evolutionary dynamics where analytical solutions may be intractable.

Here, we use an agent-based model to study the long-term evolution of the actomyosin nonlinearity, $\alpha$, under competition and limited resources. 
We interpret $\alpha$ as an individual trait subject to random fluctuations and selection through competition.
First, we demonstrate that without mutations, environmental conditions sensitively select specialized nonlinearities.
Introducing mutations removes this ecological dependence, funneling populations toward a robust evolutionary attractor where simulations converge to the biologically observed $\alpha^* \approx 4$ across diverse conditions. 
We further show that mutation rate mediates a tradeoff between robustness to short-term fluctuations in trait space and adaptability to long-term trends, shaping both agent survivability and population-level convergence. 
Together, these results reveal how ecological constraints and evolutionary tradeoffs drive convergence toward a conserved muscle nonlinearity, proposing a mechanistic explanation for the ubiquity of $\alpha^* \approx 4$ in naturally occurring muscle.

\section{Results}
\label{sec:results}

Across scales --- from microscale biochemical models of actin--myosin interactions to macroscopic phenomenological descriptions of muscle mechanics --- muscle contraction exhibits a fundamental nonlinearity in force generation.
At the molecular scale, this nonlinearity arises from velocity-dependent detachment of myosin heads \cite{mcgrath_microscale_2025, seow_hills_2013, piazzesi_skeletal_2007}.
At the macroscopic scale, it manifests as the characteristic concave force–velocity relationship, in which increasing shortening velocity reduces muscle tension \cite{hill_heat_1938}. 
We parameterize the strength of this shared micro-to-macro nonlinearity by $\alpha$.

In prior work \cite{mcgrath_microscale_2025}, we conducted a broad survey of 135 independent experimental measurements of muscle tissue and found that $\alpha$ consistently clusters around a characteristic value $\alpha^* = 3.85 \pm 2.32$ across evolutionarily distant species and muscle types.
We hypothesized that this conserved value reflects a tradeoff between power output and energetic efficiency during contraction \cite{mcgrath_microscale_2025}.
However, that study focused on short-timescale contractile performance over the duration of a mechanical task and did not consider how $\alpha$ might evolve under long-term selection pressures arising from competition and limited resources.

Here, we extend this framework by embedding the actin--myosin nonlinearity into an evolutionary setting.
We construct an agent-based simulation in which each agent actuates through its environment as a minimal $\alpha$-valued muscle and, over thousands of generations, fitness emerges from competition for finite resources.
This approach allows us to extend molecular-scale nonlinearities to influence population-level phenotypic distributions over evolutionary timescales.

\subsection{Agent-based modeling of the evolution of actin--myosin nonlinearities}
\label{sec:problem_definition}

We build on our previously developed minimal two-state model of actin--myosin dynamics \cite{mcgrath_microscale_2025}, in which myosin heads transition between attached and detached states while consuming free energy to perform mechanical work (Methods).
The resulting nondimensionalized system of equations captures the coupling between attachment dynamics, contraction velocity, and energetic cost.
We derived the following governing equations in Ref. \cite{mcgrath_microscale_2025}:
\begin{align}
    \dot{A} &= 1 - A \left(1 + \alpha V + \gamma \right), \label{eq:A} \\
    \dot{V} &= \beta A (1 - V), \label{eq:V} \\
    \dot{E} &= \zeta (1 - A) + \frac{1}{\alpha^2}. \label{eq:E}
\end{align}
where time is rescaled by the rate of constant myosin attachment, $A(t)$ denotes the fraction of myosin motors attached to actin, $V(t)$ captures the contraction velocity normalized by the muscle's no-load velocity, and $E(t)$ is a dimensionless measure of accumulated Gibbs free energy consumption.
Furthermore, there are four dimensionless parameters that characterize the system.
The first, $\alpha$, characterizes the velocity-dependent detachment of myosin from actin, thereby introducing a nonlinear coupling term between $A(t)$ and $V(t)$.
The parameter $\gamma$ represents velocity-independent detachment of myosin, $\beta$ relates muscle actuation properties to the inertia of an attached mass, and $\zeta$ relates Gibbs free energy consumption to a characteristic mechanical power.
As shown in Ref. \cite{mcgrath_microscale_2025}, a simple calculation demonstrates that $\gamma$, $\beta$, and $\zeta$ are all of order one; we therefore fix them to 1 in all simulations, allowing only $\alpha$ to vary within its biologically observed range ($\alpha > 0$).

\begin{figure}[h]
    \centering
    \includegraphics[width=0.85\linewidth]{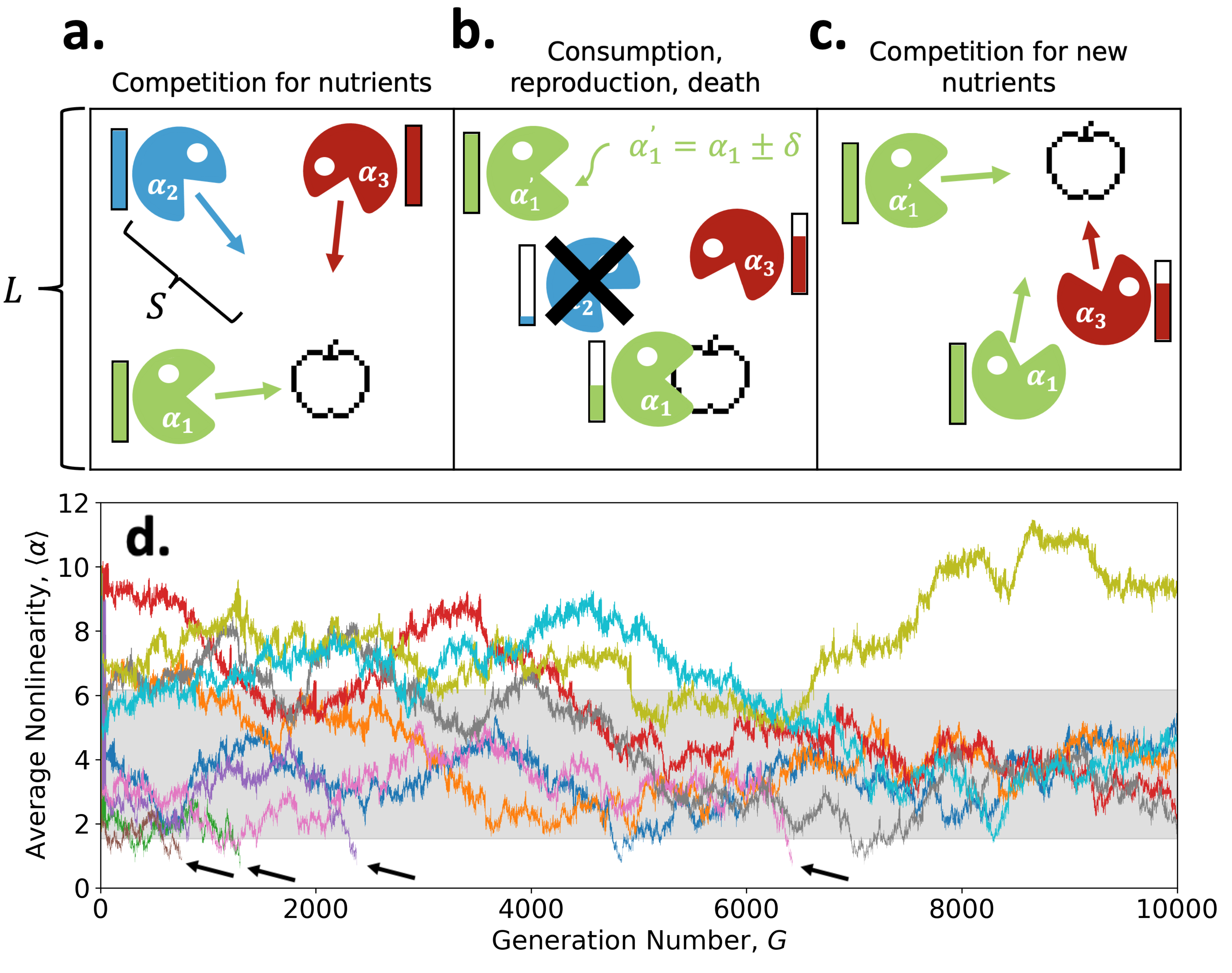}
    \captionsetup{font=footnotesize}
    \caption{Agent-based model of competing minimal-muscle agents. 
        (a) $n$ minimal-muscle agents with unique $\alpha$ values are randomly initialized on an $L \times L$ periodic domain with finite energy reserves. $S$ denotes the agents' range --- defined as the distance an $\alpha=1$ agent can travel before depletion. A nutrient is placed at a random location, and agents actuate toward it according to Eqs.~\ref{eq:A}--\ref{eq:E}. 
        (b) Agents that exhaust their reserves die and are removed. The first agent to reach the nutrient consumes it, restores its energy reserves, and produces an offspring with mutated trait $\alpha_i'=\alpha_i \pm \delta$, where $\delta \sim \mathcal{N}(0,\delta^2)$. 
        (c) The nutrient is repositioned randomly, and the process repeats for $G$ generations, enabling selection. 
        (d) Representative trajectories of the population mean $\langle \alpha \rangle$ for $N=10$ trials over $10^4$ generations. The light gray band denotes the region around $\alpha^* \approx 4$. Five trials converge near $\alpha^*$, one evolves toward $\langle \alpha \rangle > \alpha^*$, and four terminate in extinction (arrows).}
    \label{fig:setup}
\end{figure}

To investigate the evolutionary convergence of muscle force–velocity curvature, $\alpha$, under competition and limited resources, we develop an agent-based simulation that incorporates the actomyosin dynamics above (Methods).
We consider an $L \times L$ periodic domain populated with $n$ agents initialized at random positions as in Fig.~\ref{fig:setup}a.
Each agent is assigned a unique $\alpha$ value and a finite initial internal energy reserve $E(0)$.
Each agent converts stored internal energy reserves through $\dot{E}$ into mechanical work through $\dot{A}$ and $\dot{V}$, effectively behaving as a minimal muscle-like actuator navigating a resource landscape.
Next, to introduce competition, a nutrient resource is placed at a random location in the domain.
Agents expend energy to move toward it, as in Fig.~\ref{fig:setup}b, and the first agent to reach the nutrient consumes it and fully replenishes its reserves.
Additionally, this agent produces an offspring with a mutated trait $\alpha'=\alpha \pm \delta$ (i.e., reproduction, Fig.~\ref{fig:setup}b), where $\delta$ is drawn from a normal distribution $\mathcal{N}(0,\delta^2)$.
Then, upon consumption, the agents are randomly displaced from their final positions to avoid clustering around the nutrient.
Meanwhile, agents that fully deplete their reserves are removed from the population (i.e., death, Fig.~\ref{fig:setup}b).
A new nutrient source is repositioned to a random location, and the competition repeats (Fig.~\ref{fig:setup}c).
This cycle --- resource placement, competition and death, nourishment and reproduction --- is repeated over many generations.
Natural selection thus acts on the trait $\alpha$ and produces population-level distributions $P(\alpha)$ that evolve over generations.

To characterize the effective resource availability of the environment, we quantify each agent’s energy supply through a distance scale $S$, which we term the "range", defined as the distance an $\alpha = 1$ agent can travel before fully depleting its reserves.
Because agents convert stored energy into actuation according to Eqs.~\ref{eq:A}--\ref{eq:E}, the effective range of an agent depends on $\alpha$: higher-$\alpha$ agents expend energy more slowly (i.e., $\dot{E}$ grows more gradually), and therefore travel farther for a fixed initial reserve $E(0)$.
With $S$ denoting the range of an $\alpha = 1$ agent (set by $E(0)$) and $L$ the system size, the dimensionless ratio $S/L$ sets the effective resource availability.
Small $S/L$ corresponds to energy-limited environments, whereas large $S/L$ characterizes abundance.

Fig.~\ref{fig:setup}d displays the evolution of a population's mean trait $\langle \alpha \rangle$ over $10^4$ generations for ten representative trials.
We observe that $\langle \alpha \rangle$ exhibits a diffusive-like trajectory in trait space across values spanning $\alpha \sim 1$--$12$.
In many trials, the population's $\langle \alpha \rangle$ relaxes to a narrow band centered around the biologically relevant $\alpha^*$ (Fig.~\ref{fig:setup}d, light gray region) after $10^4$ generations, which we interpret to be steady-state.
We additionally observe extinction events (Fig.~\ref{fig:setup}d, black arrows), in which all agents within a trial die.
One trial (green trajectory) drifts beyond the gray band and explores a niche region of trait space with $\langle \alpha \rangle > \alpha^*$.

Because each simulation is initialized with a uniform distribution of $\alpha$ between 1--10, the repeated convergence of independent lineages toward the narrow band centered at $\alpha^* \approx 4$ suggests that this trait confers a selective advantage.
At the same time, extinction events and occasional excursions away from $\alpha^*$ indicate that the dynamics remain stochastic and sensitive to demographic and environmental fluctuations.
Nevertheless, the persistent relaxation toward $\alpha^*$ despite inter-generational noise leads us to hypothesize that $\alpha^*$ represents a robust evolutionary attractor of the underlying actomyosin dynamics, whose stability depends on energy availability $S/L$ and mutation rate $\delta$.

\subsection{Adaption overcomes scarcity but can destabilize populations}
\label{sec:reproduction}

To disentangle the concurrent roles of mutation rate $\delta$ and resource availability $S/L$ in the evolution of our agent-based model, we first examine the dynamics without mutations by setting $\delta = 0$ across three levels of resource availability, $S/L = \sqrt{2} \cdot (\frac{1}{2}, \frac{3}{4}, \frac{5}{4})$. 
The $\sqrt{2}$ prefactor accounts for the maximum possible distance an agent can traverse in the $L \times L$ domain, so writing $S/L = \sqrt{2} \cdot x$ indicates that an $\alpha=1$ agent can cover a fraction $x$ of the longest possible distance in the domain before depletion. 
The chosen values span conditions from energy-limited environments ($S/L = \frac{\sqrt{2}}{2}$, where $\alpha=1$ agents can only traverse half of the domain before depletion) to energy-abundant environments ($S/L = \frac{5\sqrt{2}}{4}$, where $\alpha=1$ agents have sufficient energy to travel farther than the maximum possible distance in the domain).
We do not extend beyond $S/L > \frac{5\sqrt{2}}{4}$, as extreme energy abundance leads to explosive population growth and computationally prohibitive simulations. 

\begin{figure}[h]
    \centering
    \includegraphics[width=0.85\linewidth]{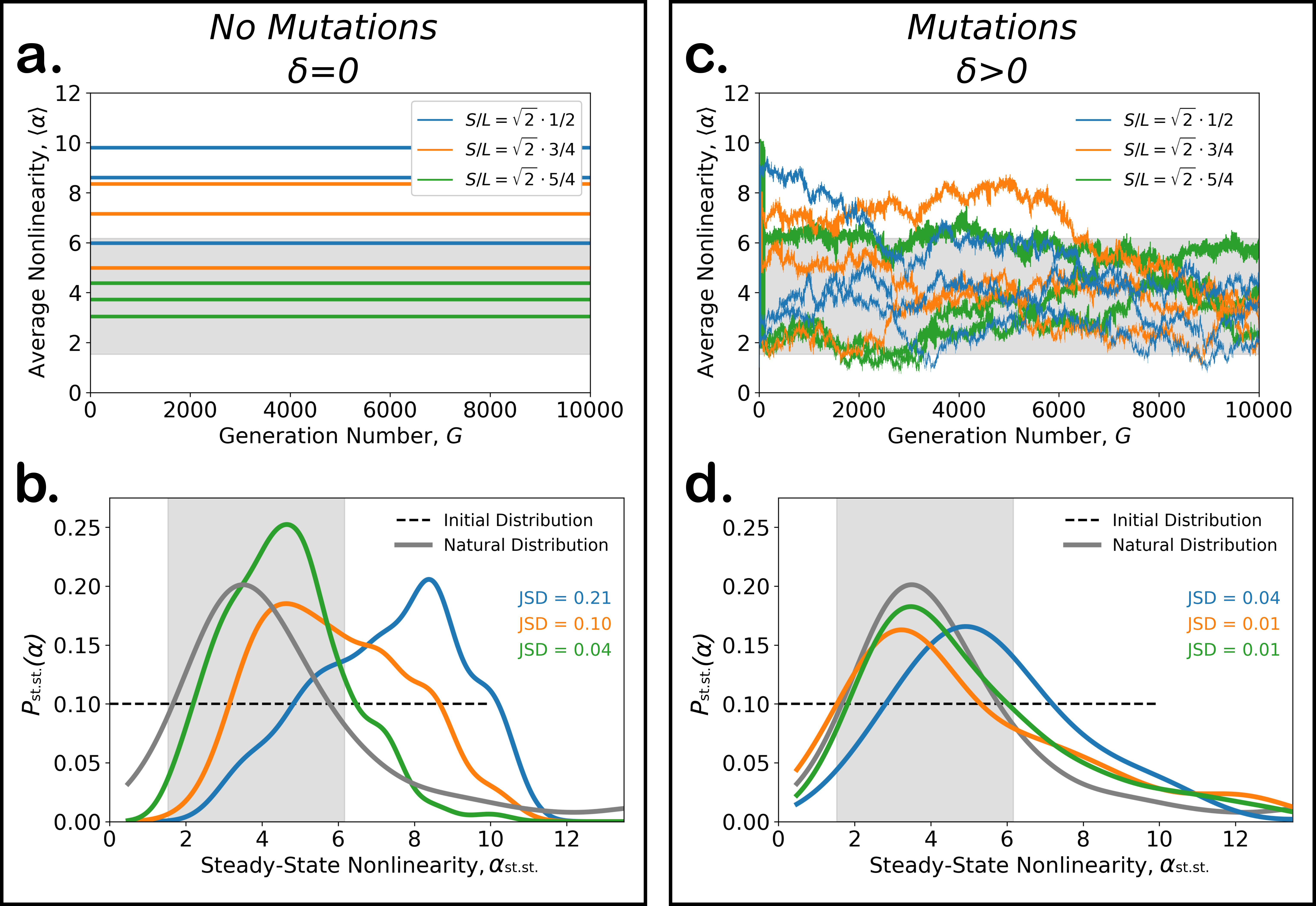}
    \captionsetup{font=footnotesize}
    \caption{Adaptation and selection together drive evolutionary convergence toward $\alpha^*$. 
        (a) Mean trait $\langle \alpha \rangle$ trajectories without mutations ($\delta = 0$) across three values of the resource level $S/L$ (color corresponds to value, see legend). 
        For each value of $S/L$, three representative trials are shown. 
        (b) Over many trials, the resulting steady-state distribution of $\alpha$, $P_{\rm st.st.}(\alpha)$, is unimodal (curve color corresponding to legend in panel (a)), demonstrating evolutionary convergence. 
        However, the distributions do not resemble the natural distribution $P^*(\alpha)$ (gray curve).
        (c) Mean trait trajectories for three representative trials with mutations ($\delta > 0$) for the same $S/L$ levels. 
        (d) With mutations, populations converge toward $\alpha^* \approx 4$ (light gray band) regardless of $S/L$.
        The Jensen–Shannon divergence $\mathrm{JSD}(P_{\rm st.st.}\|P^*)$ decreases for all $S/L$ when mutations are introduced (comparing panels (b) and (d)), quantifying convergence toward $\alpha^*$. 
        Dashed lines in (b,d) indicate the initial uniform trait distribution at $t=0$.}
    \label{fig:with_and_without_mutations}
\end{figure}

Fig.~\ref{fig:with_and_without_mutations}a shows the trajectory of the population mean $\langle \alpha \rangle$ for three representative trials across each tested level of $S/L$ where each population’s mean trait $\langle \alpha \rangle$ remains frozen, as mutations cannot explore $\alpha$-space when $\delta = 0$. 
For each tested level of resource availability $S/L = \sqrt{2} \cdot (\frac{1}{2}, \frac{3}{4}, \frac{5}{4})$ (blue, orange, green curves), we run 650, 750, and 500 independent simulations, respectively, to estimate the steady-state distribution $P_{\rm st.st.}(\alpha) = \lim_{t\rightarrow \infty}P(\alpha,t)$. 
Starting from a uniform initial distribution $P(\alpha,t=0)$ (Fig.~\ref{fig:with_and_without_mutations}b, black dashed line), the trait's distribution in steady-state $P_{\rm st.st.}(\alpha)$ is approximately unimodal, demonstrating that selection favors particular $\alpha$-values. 
Moreover, as $S/L$ decreases without mutations, the peak of $P_{\rm st.st.}(\alpha)$ increases, indicating that selection favors more energy-conserving, specialized traits under energy-limited conditions.

Next, we examine the effect of allowing selection to explore trait space by introducing a finite mutation rate, $\delta > 0$. 
Fig.~\ref{fig:with_and_without_mutations}c shows the evolution of the population mean trait $\langle \alpha \rangle$ for three representative trials at each level of resource availability, $S/L = \sqrt{2} \cdot (\frac{1}{2}, \frac{3}{4}, \frac{5}{4})$. 
Lineages fluctuate stochastically due to mutations.
However, over long timescales these fluctuations give way to a systematic convergence toward $\alpha^*$. 
Repeating these in silico experiments 150 times per $S/L$ level, we construct steady-state trait distributions $P_{\rm st.st.}(\alpha)$ in Fig.~\ref{fig:with_and_without_mutations}d.
Introducing mutations leads to evolutionary convergence toward $\alpha^*$, as evidenced by unimodal distributions centered around this value. 
Crucially, the location of the peak for each $P_{\rm st.st.}(\alpha)$ --- the most likely trait for a population to converge upon --- is largely insensitive to $S/L$, in stark contrast to the no mutations $\delta = 0$ case in panel b.

Furthermore, the steady-state trait distributions $P_{\rm st.st.}(\alpha)$ with mutations ($\delta > 0$) more closely resemble the naturally occurring distribution $P^*(\alpha)$ (gray distributions, Fig.~\ref{fig:with_and_without_mutations}b,d) than those obtained without mutations ($\delta = 0$). 
To quantify this similarity, we compute the Jensen-Shannon divergence (a symmetric variant of the Kullback–Leibler divergence, Methods), $\mathrm{JSD}(P_{\rm st.st.}\|P^*)$, between each distribution obtained from simulation $P_{\rm st.st.}$ and the natural distribution $P^*$, where lower values of $\mathrm{JSD}$ indicate greater similarity between the two distributions. 
From Fig.~\ref{fig:with_and_without_mutations}b to Fig.~\ref{fig:with_and_without_mutations}d, $\mathrm{JSD}(P_{\rm st.st.}\|P^*)$ decreases across all tested $S/L$ levels, showing that introducing mutations drives the steady-state distributions closer to the natural distribution where $\left<\alpha\right> \approx \alpha^* \approx 4$.

In the absence of mutations ($\delta = 0$), long-term survival is primarily determined by the dimensionless resource availability $S/L$. 
When $S/L$ is small, the environment is energy-limited: agents traverse long distances relative to their energy reserves, favoring energy-efficient (high-$\alpha$) strategies. 
Conversely, when $S/L$ is large, nutrients are abundant, and powerful, low-$\alpha$ agents can move quickly to their target without significant energetic constraints. 
These results show that, without mutation, selection acts primarily through environmental filtering, and populations do not generically converge to the biologically observed trait $\alpha^*$. 
Instead, survival outcomes are dictated by energetic constraints rather than an intrinsic optimality in actin--myosin nonlinearities.
Introducing mutations ($\delta > 0$) fundamentally changes this picture. 
Adaptive dynamics allow lineages to explore trait space, enabling selection to act on variations in $\langle \alpha \rangle$. 
Populations systematically settle to $\alpha^*$, producing steady-state distributions $P_{\rm st.st.}(\alpha)$ that are largely independent of $S/L$ and closely match the naturally observed distribution. 
This convergence reflects a selective advantage around $\alpha^*$: when an efficient parent ($\alpha > \alpha^*$) gives birth to offspring with $\alpha \pm \delta$, those that retain high efficiency while gaining additional power (slightly lower $\alpha$) are favored. 
And conversely, the slightly more efficient offspring of powerful parents appear to be favored. 
These complementary biases establish a stable evolutionary attractor at $\alpha^*$ and drive population-level distributions toward this characteristic value.
Together, these results demonstrate that environmental constraints can filter the feasible trait space, but evolutionary exploration via mutations and selection drives convergence to the biologically observed $\alpha^*$. 

Evolutionary success, however, depends not only on convergence through adaptation but also on the stability of that adaptation. 
The mutation rate $\delta$ determines how rapidly populations explore trait space, regulating the balance between responsiveness to selection pressures and robustness to stochastic fluctuations in $\langle \alpha \rangle$. 
And when the mutation rate is too large, deleterious mutations can overwhelm adaptation \cite{sprouffske_high_2018}.
We therefore next investigate how the mutation rate $\delta$ affects the stability of evolving populations across a range of resource availabilities $S/L$.

\begin{figure}[h]
    \centering
    \includegraphics[width=0.6\linewidth]{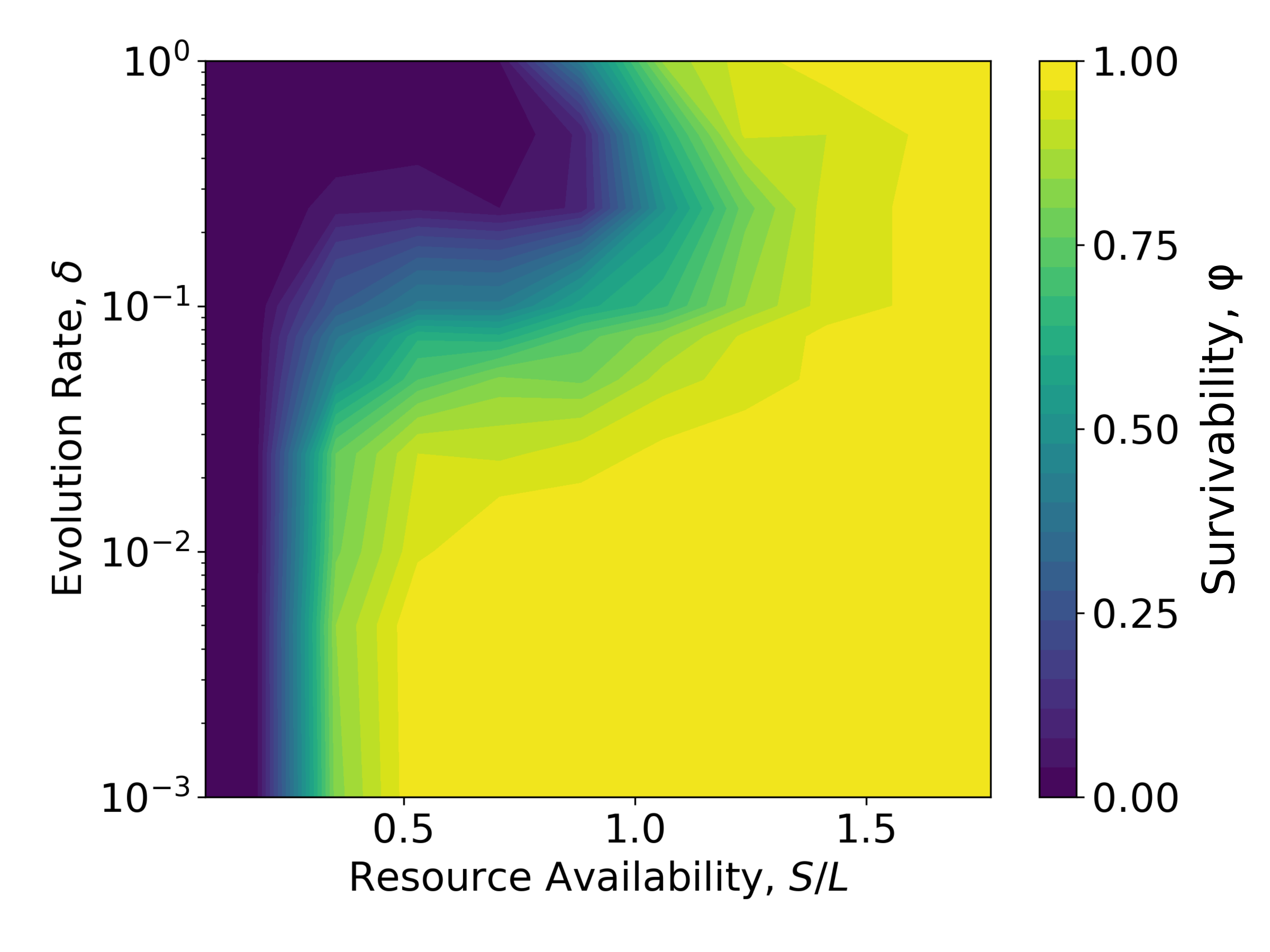}
    \captionsetup{font=footnotesize}
    \caption{The population survivability, $\phi$, (i.e. the fraction of trials that avoid extinction) is governed by the interplay between mutation rate $\delta$ and resource availability $S/L$. 
    Low resource availability and/or high mutation rates destabilize populations and increase extinction probability (dark regions). 
    In contrast, sufficient resources combined with moderate or low mutation rates promote stable, persistent populations (light regions).}
    \label{fig:extinctions}
\end{figure}

For 100 distinct $(S/L, \delta)$ pairings, with mutation rates spanning $10^{-3}$--$10^{0}$ and resource availabilities spanning $\sqrt{2}/20$--$5\sqrt{2}/4$, we conducted at least $150$ independent trials evolved for $10^4$ generations. 
For each $(S/L, \delta)$ combination, we compute the fraction of trials that avoid extinction (which we term survivability, $\phi$, shown in Fig.~\ref{fig:extinctions}).

We identify a critical resource threshold near $S/L \approx \sqrt{2}/4$ in Fig.~\ref{fig:extinctions}, below which extinction is inevitable. 
In this regime, the domain size $L$ is large relative to the agents’ energy reserves $S$, causing individuals to exhaust their energy before reaching nutrients. 
The energetic deficit is sufficiently severe that no mutation rate can generate adaptive responses quickly enough to rescue the population, and collapse occurs across all $\delta$. 
By contrast, when $S/L \gtrsim \sqrt{2}/4$, nutrients are accessible and sustained survival becomes possible.

Within this survivable regime, mutation rate modulates stability. 
When $\delta$ becomes large ($\delta \gtrsim 10^{-1}$), survivability declines for an extended range of resource availabilities. 
This decline suggests that rapid phenotypic exploration can destabilize evolving populations, potentially because large fluctuations in $\langle \alpha \rangle$ prevents populations from reliably tracking selective pressures. 
Large mutations can transiently favor extreme, energetically inefficient low-$\alpha$ strategies, leading to rapid energy depletion and extinction. 
Consequently, a substantial fraction of simulations collapse at high $\delta$ unless resources are sufficiently abundant ($S/L \gtrsim \sqrt{2}$) to buffer these stochastic excursions.
Thus, while mutation is necessary for adaptive convergence, excessively rapid evolution can destabilize populations in resource-limited environments. 
Stable evolutionary outcomes require sufficient energy availability and moderate rates of phenotypic exploration.

\subsection{Convergent outcomes, and their predictability, are shaped by resource availability and mutation rate}
\label{sec:convergence}

Previous in silico experiments revealed two key features of the model.
In the absence of mutations ($\delta = 0$), evolutionary outcomes are primarily determined by resource availability.
However, when mutations are introduced ($\delta > 0$), selection drives adaptive exploration in trait space toward $\alpha^*$.
At the same time, excessively large mutation rates ($\delta \gtrsim 10^{-1}$) can destabilize populations, particularly in resource-limited environments.
These results raise a broader question: beyond simple survival, how do the competing influences of environmental constraints ($S/L$) and evolutionary exploration ($\delta$) shape the steady-state trait distributions $P_{\rm st.st.}(\alpha)$?

\begin{figure}[h]
    \centering
    \includegraphics[width=0.9\linewidth]{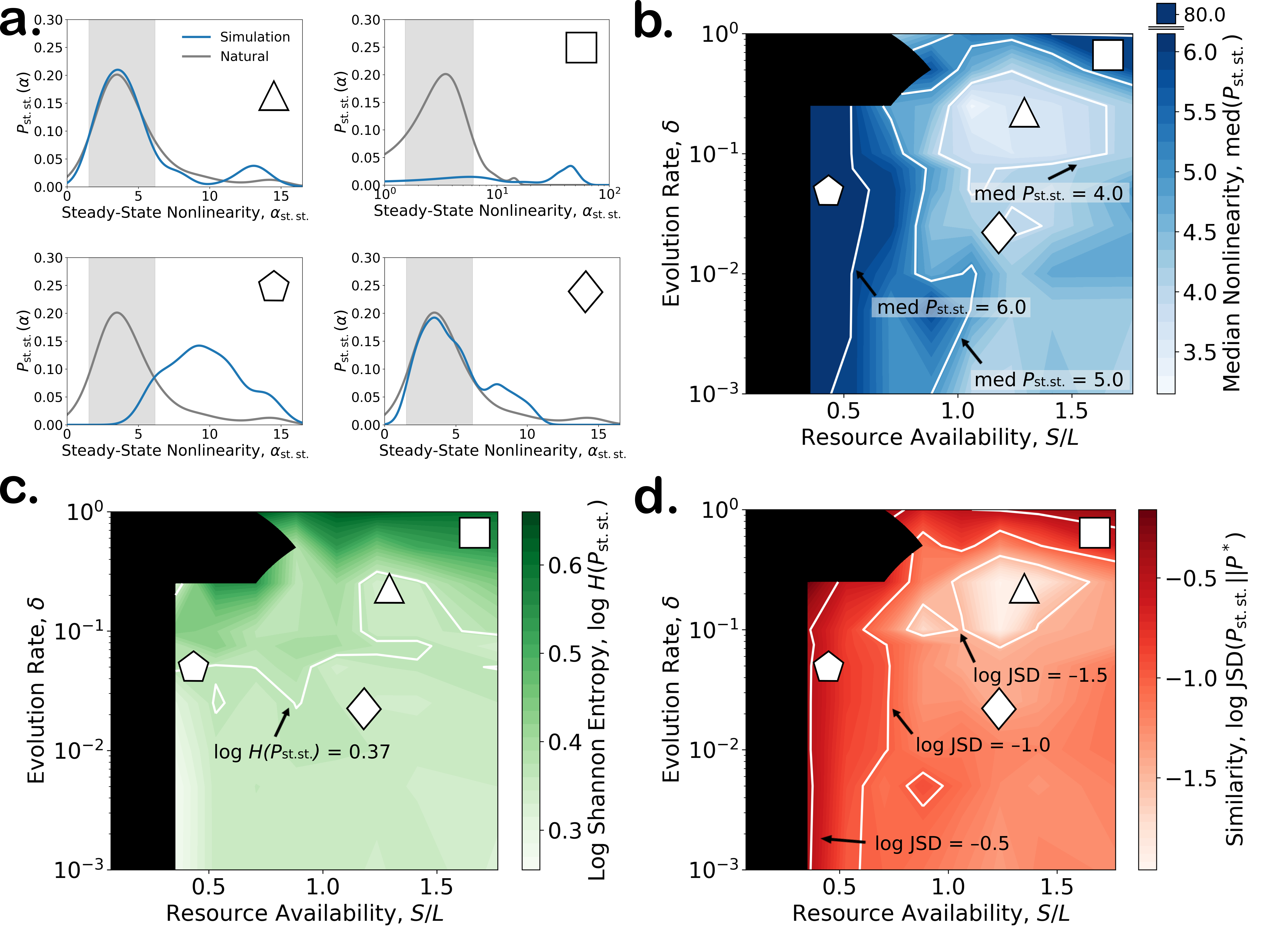}
    \captionsetup{font=footnotesize}
    \caption{Resource availability $S/L$ and mutation rate $\delta$ shape convergence outcomes of $P_{\rm st. st.}(\alpha)$.
        (a) Four representative distributions (blue) for distinct $(S/L, \delta)$ conditions, compared with the natural distribution (gray).
        Triangle ($S/L,\delta = \sqrt2,1/4$), square ($S/L,\delta = 5\sqrt2/4,1$), pentagon ($S/L,\delta = \sqrt2/4,1/20$), and diamond ($S/L,\delta = 7\sqrt2/8,1/40$) indicate the locations of these distributions in the phase-space plots in panels (b–d).
        (b) Typical evolutionary outcome. 
        Median evolved steady-state nonlinearity, $\mathrm{med}(P_{\rm st.st.})$, across all tested $(S/L, \delta)$ combinations. 
        Contours indicate $\mathrm{med}(P_{\rm st.st.}) = 4, 5, 6$. 
        The color-bar lookup table is limited to 6.17 for readability. 
        Extreme nonlinearities arise in resource-limited environments or at high mutation rates, whereas moderate mutation rates in abundant environments produce values consistent with the natural distribution.
        Evolved outcomes are strongly dictated by resource availability in the low mutation limit.
        (c) Predictability of convergence. 
        Logarithm of the Shannon entropy, $\log(H)$, computed from the simulated $P_{\rm st.st.}$ distributions, quantifies uncertainty in evolutionary outcomes. 
        Contour indicates $\log(H) = 0.37$, the value calculated from the natural distribution. 
        Lower mutation rates yield more deterministic outcomes; higher rates increase uncertainty in evolving trajectories.
        (d) Similarity to the natural distribution. 
        Logarithm of the Jensen–Shannon divergence, $\log(\mathrm{JSD}(P_{\rm st.st.}\|P^*))$, quantifies how closely each simulated distribution matches the natural distribution. 
        Contours denote $\log(\mathrm{JSD}) = -1.5, -1.0, -0.5$. 
        High mutation rates or low resources produce distributions that deviate strongly from nature, whereas moderate mutation rates in abundant environments yield distributions nearly identical to the natural one.}
    \label{fig:outcomes}
\end{figure}

To address this, we systematically explore evolutionary outcomes across the same 100 distinct $(S/L,\delta)$ pairings shown in Fig.~\ref{fig:extinctions}, conducting at least $150$ independent trials for each pairing evolved for $10^4$ generations.
From these simulations we construct probability distributions of the evolved steady-state nonlinearity, $P_{\rm st.st.}(\alpha)$, across the full range of environmental constraints and mutation rates.
Fig.~\ref{fig:outcomes}a shows four representative $P_{\rm st.st.}(\alpha)$ distributions for selected $(S/L,\delta)$ pairings, ranging from abundant resources with rapid mutations (square, $S/L,\delta = 5\sqrt2/4,1$) to scarcity with weak mutations (pentagon, $S/L,\delta = \sqrt2/4,1/20$).
Either extreme mutation rates or severely resource-limited environments (square and pentagon, respectively) push evolving populations toward highly specialized, niche $\alpha_{\rm st.st.}$ values, as evidenced by their peaks in $P_{\rm st.st.}(\alpha)$ shifted rightward of $\alpha^*$.
Under extreme mutation rates, we additionally observe evolved traits far outside the range observed in nature, reaching $\alpha_{\rm st.st.} \approx 80 \gg \alpha^*$.
By contrast, intermediate mutation rates in sufficiently abundant environments (triangle, $S/L,\delta = \sqrt2,1/4$; diamond, $S/L,\delta = 7\sqrt2/8,1/40$) produce evolved distributions that closely match the natural distribution (gray curves). 

To characterize evolutionary outcomes across all tested $(S/L,\delta)$ pairings, we address three questions in Fig.~\ref{fig:outcomes}:
(1) What $\alpha_{\rm st.st.}$ value does a typical lineage evolve to (Fig.~\ref{fig:outcomes}b)?
(2) With what degree of certainty does a lineage reach that value (Fig.~\ref{fig:outcomes}c)?
(3) How closely do the aggregate results match the observed natural distribution (Fig.~\ref{fig:outcomes}d)?

In Fig.~\ref{fig:outcomes}b we show the median steady-state nonlinearity, $\mathrm{med}(P_{\rm st.st.})$, as a measure of the typical evolutionary outcome for each $(S/L,\delta)$ pairing.
Both resource-limited environments and excessively high mutation rates drive the system toward extreme nonlinearities, though for different reasons.
In resource-limited environments, high-$\alpha$ agents are favored because they are energetically efficient.
In contrast, when mutation rates are large, continual exploration of trait space causes $\langle \alpha \rangle$ to fluctuate widely, occasionally pushing populations toward extreme low-$\alpha$ strategies (often leading to extinction, as in Fig.~\ref{fig:extinctions}) or toward very large $\alpha$ values (dark blue regions in Fig.~\ref{fig:outcomes}b).
Some evolved values were so extreme (median $\mathrm{med}(P_{\rm st.st.}) \approx 80$) that we capped the plot at $6.17$, corresponding to the upper quartile of the natural distribution $P^*$, to improve readability.

Across a broad region of parameter space, however, typical lineages relax toward the characteristic value $\alpha^* \approx 4$.
In particular, for moderate mutation rates ($10^{-1} \gtrsim \delta \gtrsim 10^{-2}$) and sufficiently abundant resources ($S/L \gtrsim 3\sqrt2/4$), the resulting distributions cluster around $\alpha^*$ (light blue region in Fig.~\ref{fig:outcomes}b).
By contrast, when mutations are negligible ($\delta \lesssim 10^{-2}$), the evolved median $\mathrm{med}(P_{\rm st.st.})$ closely tracks the level of resource availability, again highlighting that environmental filtering dominates trait selection in the absence of meaningful evolutionary exploration (as seen in Fig.~\ref{fig:with_and_without_mutations}a,b).

We next compute the Shannon entropy of each evolved steady-state distribution $P_{\rm st.st.}$ as a measure of the certainty of evolutionary outcomes and, equivalently, the degree of evolutionary convergence. 
In Fig.~\ref{fig:outcomes}c, we evaluate this entropy across all tested $(S/L,\delta)$ pairings. 
Shannon entropy quantifies uncertainty in a probability distribution by measuring how broadly outcomes are distributed: sharply peaked distributions that are consistently reproduced across simulations --- analogous to convergent evolution --- yield low entropy, whereas broader distributions produce higher entropy. 
Consistent with Fig.~\ref{fig:with_and_without_mutations}a,b, low mutation rates generate relatively peaked distributions, as limited evolutionary exploration allows resource availability to strongly constrain the selected steady-state nonlinearity $\alpha_{\rm st.st.}$. 
As a result, low mutation rates ($\delta \lesssim 10^{-2}$) in Fig.~\ref{fig:outcomes}c produce low entropy in $P_{\rm st.st.}$.
By contrast, at high mutation rates ($\delta \gtrsim 10^{-1}$), rapid exploration of trait space makes evolutionary outcomes less predictable, producing broad, high-entropy distributions.
In this regime, evolutionary trajectories become unpredictable, as high mutation rates cause populations to continually explore trait space rather than reliably tracking the selective pressures that would otherwise drive convergence to a repeatable steady-state solution.

Finally, in Fig.~\ref{fig:outcomes}d we compute the Jensen–Shannon divergence (Methods), $\mathrm{JSD}(P_{\rm st.st.}\|P^*)$, between each simulated phenotypic distribution $P_{\rm st.st.}$ and the natural distribution $P^*$ as a measure of statistical similarity. 
The Jensen–Shannon divergence is bounded between 0 and 1, with smaller values indicating greater similarity between distributions. 
This metric therefore provides a quantitative assessment of how closely the evolved steady-state outcomes match observations from nature.
We identify a distinct region of parameter space --- moderate mutation rates ($0.05 \le \delta \le 0.25$) combined with sufficiently abundant resources ($3\sqrt{2}/4 \le S/L \le \sqrt{2}$) --- where the evolved distributions of $\alpha_{\rm st.st.}$ closely match the natural distribution (light red regions in Fig.~\ref{fig:outcomes}d).
By contrast, at high mutation rates or under severe resource scarcity, selection favors niche traits with $\alpha_{\rm st.st.} > \alpha^*$, producing distributions that deviate strongly from the natural distribution (dark red regions in Fig.~\ref{fig:outcomes}d).
When mutation rates are low, resource availability dominates selection of $P_{\rm st.st.}(\alpha)$. 
In this regime, the resulting distributions remain relatively symmetric around a selected value of $\alpha_{\rm st.st.}$ (e.g., the pentagon case in Fig.~\ref{fig:outcomes}a). 
However, the natural distribution exhibits a pronounced tail, and therefore even when the median $\alpha_{\rm st.st.}$ may align with $\alpha^*$ (as in Fig.~\ref{fig:outcomes}b), the overall distribution still differs from that observed in nature.

Together, these results demonstrate that evolutionary outcomes in our agent-based model are jointly shaped by mutation rate and resource availability. 
High mutation rates or severely resource-limited environments drive populations toward extreme or highly specialized nonlinearities, whereas intermediate mutation rates in sufficiently abundant environments reliably produce convergent evolution toward the characteristic $\alpha^*$ observed in nature. 
Mutations not only enable populations to reach these robust steady states but also introduce variability that can facilitate adaptive exploration or, when excessive, destabilize populations. 
These findings underscore the delicate balance between evolutionary dynamics and environmental constraints in determining evolved phenotypic distributions. 
Importantly, while we have characterized the resulting patterns, the specific mechanisms by which resource availability and mutation rates interact to shape the trajectory of evolution remain to be elucidated, motivating further investigation into the interplay between selection and adaptive exploration.

\subsection{Robustness-adaptability tradeoffs facilitate evolutionary outcomes}
\label{sec:mechanisms}

Having established that mutation rate and resource availability jointly shape convergent evolutionary outcomes and their predictability, we now investigate the mechanisms underlying these patterns.
In particular, we examine how populations explore trait space across generations and how stochastic fluctuations influence the resulting steady-state distributions.

To quantify how populations shift between generations, we introduce a metric analogous to a mean-squared displacement, defined at the level of population distributions.
Specifically, we compute $\mathrm{JSD}(P_i \| P_{i+\tau})$ to measure the statistical divergence between trait distributions separated by $\tau$ generations (i.e., distributions $P(\alpha)$ at generations $i$ and $i+\tau$).
Averaging over all generation pairs with the same lag $\tau$ gives the time structure function $\langle \mathrm{JSD}(P_i \| P_{i+\tau}) \rangle_i$ (Methods).
This measure captures both stability --- when distributions change little over time --- and adaptive exploration --- when distributions evolve substantially --- allowing us to systematically assess the mechanisms behind distinct evolutionary trajectories.

\begin{figure}[h]
    \centering
    \includegraphics[width=0.9\linewidth]{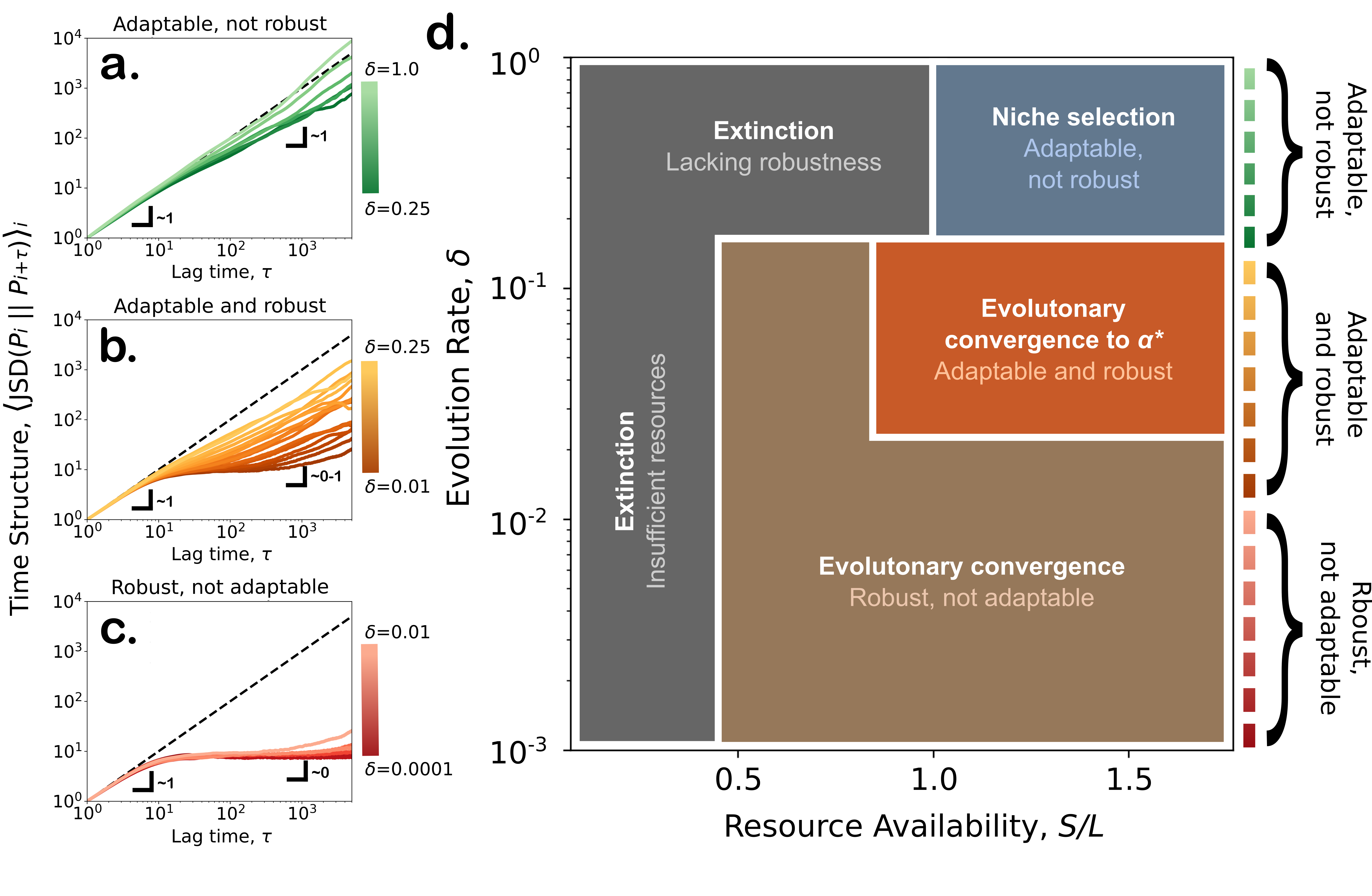}
    \captionsetup{font=footnotesize}
    \caption{Evolutionary outcomes are determined by resource availability $S/L$ and adaptation $\delta$. 
    In turn, $\delta$ sensitively determines a trade-off between robustness and adaptability, which governs evolutionary convergence. 
    (a--c) The time structure function $\langle \mathrm{JSD}(P_i \| P_{i+\tau}) \rangle_i$ measures the divergence between population distributions separated by $\tau$ generations, averaged over all $i$. 
    (a) High mutation rates ($\delta \ge 0.25$): $\alpha$ diffuses freely across generations, as evidenced by an exponent of $1$, reflecting high levels of adaptation.
    (b) Intermediate mutation rates ($0.01 \le \delta \le 0.25$): adaptation (exponent $\approx 1$) is observed at short and long timescales. 
    At intermediate timescales, the exponent decreases toward zero, indicating subdiffusive dynamics characteristic of constrained diffusion and hence evolutionary robustness.
    (c) Low mutation rates ($\delta \le 0.01$): populations are overly robust and weakly adaptable; short-term changes arise from the turnover of successive generations, but long-term transport is suppressed, resembling trapped diffusion.
    (d) Schematic illustrating distinct evolutionary regimes and the underlying mechanisms. 
    Dark grey regions: extinction events, due to limited resources or populations lacking robustness. 
    Blue region: populations are highly adaptable but not robust; traits continually explore $\alpha$-space, producing niche solutions. 
    Brown region: populations converge over repeated trials, producing robust solutions, but low mutation rates limit exploration and adaptability. 
    Here, evolved outcomes are strongly dictated by energy availability. 
    Light brown region: populations balance robustness and adaptability, convergently producing phenotypic distributions that closely match the natural distribution $P^*$. 
    }
    \label{fig:mechanisms}
\end{figure}

Using this framework, we examine how evolution rate $\delta$ mediates the trade-off between robustness and adaptability.
At high $\delta$ (Fig.~\ref{fig:mechanisms}a), distributions shift continuously across all lag times, resembling a random walk through $\alpha$-space. 
The mean $\mathrm{JSD}$ remains high with slope $\sim 1$ across all $\tau$, indicating persistent evolutionary exploration. 
Populations are highly adaptable but lack robustness, amplifying stochastic fluctuations and increasing the risk of destabilization or extinction.
Intermediate $\delta$ (Fig.~\ref{fig:mechanisms}b) produces a distinct pattern: short-term changes reflect the continual introduction of new mutants (slope $\sim 1$ for $\tau \lesssim 10$), while deviations from $\alpha^*$ are progressively suppressed at longer timescales, causing the slope to flatten ($\sim 0$–$1$). 
Here, populations achieve a balance between adaptability and robustness, and systematically converge toward the evolutionary attractor at $\alpha^*$.
At low $\delta$ (Fig.~\ref{fig:mechanisms}c), populations explore $\alpha$-space very slowly. 
The mean $\langle \mathrm{JSD}(P_i \| P_{i+\tau}) \rangle_i$ quickly plateaus, reflecting a "locked" distribution. 
Small changes occur at short timescales ($\tau \lesssim 10$) as newly mutated agents are added through reproduction or removed through depletion, producing distributional changes and drift in $\langle \mathrm{JSD}(P_i \| P_{i+\tau}) \rangle_i$ (slope $\sim 1$). 
Beyond this window ($\tau \gtrsim 10$), the population remains largely static as populations cannot explore trait space and selective pressures produce minimal shifts. 
In this regime, populations are robust but poorly adaptable.

Figure~\ref{fig:mechanisms}d schematically summarizes these dynamics: resource availability sets an energetic threshold for viability, while evolution rate tunes the trade-off between population-level adaptability and robustness. 
Too low $S/L$ and energy is scarce so populations collapse.
Above a critical $S/L$, energy is sufficient for populations to survive. 
In this energy-sufficient regime, low $\delta$ limits adaptation, creating robust, static populations.
Conversely, too high $\delta$ undermines stability, and populations randomly explore $\alpha$-space.
Intermediate $\delta$, however, produces populations that are both robust to stochastic fluctuations and responsive to environmental feedback, yielding phenotypic distributions closely matching those observed in nature.

Interpreting these results, the agent-based simulation is governed by two competing timescales that shape evolutionary success. 
On short timescales, agents compete to reach the nutrient source first, favoring high-power, fast-actuating strategies (low-$\alpha$ agents). 
On longer timescales, survival is determined by the gradual depletion of finite energy reserves, favoring slower but more energy-efficient strategies (high-$\alpha$ agents). 
The interaction between these timescales creates an inherent tension: agents must balance rapid actuation against the energetic cost required to sustain it. 
Whether speed, efficiency, or an intermediate strategy is favored depends sensitively on environmental energetic constraints and the rate at which evolution can respond to them.

In regimes with sufficient energy availability but very low evolution rates ($\delta \sim 10^{-3}-10^{-4}$), exploration of $\alpha$-space is strongly constrained. 
Each offspring $\alpha'$ differs only minimally from its parent, and selective pressures coupled with the fitness landscape limit diffusion of the population mean $\langle \alpha \rangle$. 
In the absence of adaptive shifts, populations respond primarily through selective pruning, retaining only agents suited to the current environmental energy conditions.
At the opposite extreme, high evolution rates ($\delta \sim 10^{-1}$–$10^{0}$) destabilize $\alpha$ distributions. 
Rapid phenotypic turnover amplifies stochastic fluctuations, reducing robustness to environmental constraints. 
As a result, many simulations either terminate in extinction or evolve populations toward extreme, specialized $\langle \alpha \rangle$ values.
At intermediate evolution rates, populations converge toward the biologically observed value $\alpha^*$ while minimizing extinction events. 
In this regime, the evolution rate balances adaptability to environmental pressures with robustness against stochastic fluctuations in $\alpha$, allowing efficient exploration of trait space without destabilizing population structure.

\section{Discussion}

Our results reinforce the view that evolution does not optimize a single trait \cite{gould_spandrels_1979, smith_optimization_1978}, but instead navigates a Pareto-like tradeoff \cite{li_single_2019, shoval_evolutionary_2012}: within our model, gains in energetic efficiency necessarily limit mechanical power output, and vice versa. 
This tradeoff generates competing selection pressures across timescales. 
On shorter, selection-based timescales, agents compete to reach nutrients first, favoring high-power, fast-actuating (low-$\alpha$) strategies.
On slightly longer, selection-based timescales, survival under finite energy reserves favors energy efficient (high-$\alpha$) strategies. 
The resulting tension requires agents to balance rapid actuation against sustainable energy use. 
Across still longer evolutionary timescales, the mutation rate $\delta$ regulates a separate balance between robustness and adaptability, and thereby determines the rate at which populations feed back onto the ecological and environmental pressures they experience (Fig. \ref{fig:mechanisms}). 
When feedback is too strong (Fig. \ref{fig:mechanisms}a), populations destabilize or select for extreme traits (Fig. \ref{fig:extinctions}, \ref{fig:outcomes}b); when feedback is weak (Fig. \ref{fig:mechanisms}c), selection is strongly governed by resource availability (Fig. \ref{fig:with_and_without_mutations}b, \ref{fig:outcomes}b).
Provided sufficient resources and moderate feedback, the system simultaneously maintains robustness to stochastic fluctuations and responsiveness to environmental pressures (Fig. \ref{fig:mechanisms}b) and this balance yields phenotypic distributions $P_{\rm st. st.}(\alpha)$ that closely resemble those observed in nature $P^*(\alpha)$ (Fig. \ref{fig:outcomes}d).

More broadly, our work situates itself within a large body of literature that employs agent-based models to investigate evolutionary dynamics in heterogeneous populations \cite{lamarins_importance_2022, adami_evolutionary_2016}. 
ABMs have been widely used to study the emergence of cooperation \cite{amaral_heterogeneous_2018, ichinose_adaptive_2013}, eco-evolutionary feedbacks \cite{lamarins_importance_2022}, and adaptive trait distributions in spatially structured environments \cite{yang_agent-based_2024}. 
In such contexts, ABMs are particularly valuable because they allow selection to arise from explicit mechanistic interactions between individuals and their environment, without imposing a priori assumptions about fitness landscapes or mean-field dynamics. 
This bottom-up formulation makes it possible to capture selective feedback dynamics that are often inaccessible in equation-based approaches. 

Despite their flexibility and intuitive appeal, agent-based models (ABMs) have several important limitations. 
First, ABMs can be computationally intensive. 
In our case, each generation requires $O(n)$ numerical integrations of Eqs.~\eqref{eq:A}--\eqref{eq:E} for $n$ agents, and as populations grow and simulations extend over long evolutionary timescales, systematic parameter sweeps and generating sufficient statistics become increasingly costly. 
Second, ABMs often involve complex interacting rules and stochastic elements, which can obscure causal structure \cite{lee_complexities_2015} and reproducibility \cite{daly_quo_2022}, limiting the interpretation of generalizable principles. 
They can also be sensitive to modeling choices \cite{sun_simple_2016} --- such as update schemes, boundary conditions, or stochastic implementation --- and parameter calibration \cite{quera-bofarull_2023} may be challenging, particularly when empirical data are sparse or when multiple parameter combinations yield qualitatively similar outcomes. 
In our case, our model (Eqs.~\eqref{eq:A}--\eqref{eq:E}) links microscopic dynamics directly to evolutionary-scale behavior, implicitly coarse-graining many intermediate processes and thereby requiring specific parameter choices that are not uniquely constrained. 
While emergent behaviors may be numerically robust, they often lack analytical transparency, limiting mechanistic insight \cite{an_challenges_2021}. 
For these reasons, ABMs are often most effective when complemented by reduced mathematical descriptions or continuum models that clarify the underlying mechanisms of the simulated dynamics.

Coupling these results, therefore, to an explicit mathematical framework could address these limitations and provide analytical insight into the stochastically driven results of the ABM. 
In particular, the inter-generational evolution of the $\alpha$ distributions could be modeled using a Fokker--Planck equation, which describes the time evolution of probability densities. 
This approach would allow perturbative analysis around steady states $P_{\rm st.st.}(\alpha)$ and provide a continuum description of the stochastic dynamics. 
Characterizing the effective convergent forcing and diffusion terms in the Fokker--Planck formulation could further clarify the nature of the evolutionary attractor at $\alpha^*$ and further identify the mechanisms governing stability and convergence beyond that discussed in this manuscript.

As a natural extension of this work, one could dynamically tune the resource availability $S/L$ and mutation rate $\delta$ to drive transitions between distinct regimes of phenotypic distributions in $\alpha$. 
Treating these parameters as external inputs reframes the agent-based model as a feedback system, consistent with the view that biological systems can be interpreted through a control-theoretic lens \cite{an_optimization_2017, clarke_control_2025, alvarado_optimal_2026}. 
This perspective would allow characterization of response time, gain, and stability as the population adapts to perturbations in $S/L, \delta$.
The minimal ABM could also be extended to include spatial-temporal hazards. 
Spatial heterogeneities --- such as inaccessible regions within the $L \times L$ domain or areas of reduced resource density --- and temporal variability --- such as periodic or stochastic "drought" events that transiently reduce $S/L$ --- would enable systematic tests of resilience under nonstationary conditions.
Finally, incorporating predator--prey interactions or mobile nutrient fields would relax the assumption of stationary resources and introduce additional selection pressures coupled to $\alpha$. 
These extensions would probe whether the emergent distributions in $\alpha^*$ persist under greater, and more realistic, spatial-temporal complexity and pressures.

More broadly, our results suggest parallels between evolutionary adaptability and recent advances in learning and self-tuning in physical systems, including soft, active, and cytoskeletal systems. 
Local, iterative rules can endow networks with functional plasticity: spring networks can physically-learn and adapt through local rules without external computation \cite{altman_experimental_2024}, and nonlinear analog circuits rapidly learn tasks that linear systems cannot \cite{dillavou_machine_2024}. 
In biological matter, actin networks maintain mechanical stability through continuous reorganization \cite{wang_mechanosensitive_2025}, while elastic materials can be driven to highly adaptable states by alternating training objectives \cite{falk_learning_2023}. 
Insights from load-dependent actomyosin dynamics \cite{sakamoto_crosslinked_2026} and cellular control strategies for size regulation \cite{rizzo_mechanochemical_2024} further suggest that local learning and self-tuning may represent a general strategy for achieving robust, adaptive behavior across physical systems. 
We anticipate that further research at the intersection of evolutionary dynamics, learning, adaptation, and active matter will inform the development of novel materials that autonomously adapt to their mechanical environment \cite{Dong-2024-MRSBulletin}.

\section{Conclusions}
Our results demonstrate that the evolution of muscle-like nonlinearities $\alpha$ in an agent-based model is governed by the interplay between resource availability and mutation-driven adaptation. 
In the absence of mutation, selection favors strategies dictated by energetic constraints: high-$\alpha$ agents prevail in energy-limited environments, whereas low-$\alpha$ agents dominate in energy-surplus environments, and the characteristic biological value $\alpha^* \approx 4$ does not generally emerge. 
Introducing mutations enables populations to explore phenotypic space driven by selection, producing a systematic convergence toward $\alpha^*$ across a wide range of conditions. 
However, the mutation rate $\delta$ critically modulates this process: low $\delta$ produces overly robust populations that adapt slowly, high $\delta$ destabilizes populations through stochastic excursions and increases extinction risk, while intermediate $\delta$, with sufficient resources, navigates a delicate balance between robustness to short-term noise and adaptability to long-term trends, allowing populations to converge reliably to $\alpha^*$. 
Our results highlight how molecular-scale actomyosin nonlinearities are shaped by environmental constraints and evolutionary pressures over evolutionary timescales to population-level phenotypic distributions, providing a mechanism that may underlie the ubiquity of $\alpha^* \approx 4$ in nature.
Leveraging these principles could lead to the rational design of a new class of autonomous, adaptive materials.

\section{Methods}
\label{sec:methods}
\subsection{The 2 state actin--myosin model}

We refer the reader to \cite{mcgrath_microscale_2025} for a full derivation of the actomyosin dynamics used in our agent-based model. 
In what follows, we briefly summarize the governing equations implemented in the agent-based model (ABM) which captures an agent's conversion of stored energy to actuation.

We begin with a simple two-state model of actin--myosin interactions in which a fraction $A$ of myosin motors are attached to actin and a fraction $D$ are detached, with $A + D = 1$. 
We assume a constant attachment rate $k_a$, while detachment occurs at a velocity-dependent rate
\[
k_d(V) = K_D V + k_0,
\]
for some finite isometric detachment rate $k_0$. 
Rescaling time by myosin's attachment rate $\hat{t} = t k_a$ gives the normalized attachment dynamics
\[
\dot{A} = 1 - A(1 + \alpha V + \gamma),
\]
where $\alpha = K_D/k_a$ and $\gamma = k_0/k_a$. 

To couple an agent's binding dynamics to actuation within the ABM, we consider contraction of a mass $m$ driven by the ensemble of myosin motors. 
The force generated by each agent, effectively modeled as a minimal muscle, is proportional to the fraction of attached motors $A$ and the linear force--velocity relation of an individual myosin motor $F \sim 1-V$. 
Rescaling time again by $k_a$ and normalizing velocity by the motor’s maximum shortening velocity $v_0$ yields agent acceleration
\[
\dot{V} = \beta A(1 - V),
\]
where $\beta = \frac{f_0}{m v_0 k_a}$.

We extend the model to include energetic costs associated with ATP consumption. 
Myosin hydrolysis releases Gibbs free energy $g_0$, and since the total number of attaching motors is $N k_a(1-A)$, an agent's rate of energy consumption is $\dot{g} = g_0 N k_a(1-A)$. 
Experiments further demonstrate the rate of maintenance heat in muscle is $\dot{h}_m = ab$, where $a$ and $b$ are the constants from Hill's muscle model. 
Again rescaling time by $\hat{t}=t k_a$ and normalizing by a characteristic power of the ensemble of motors, $N f_0 v_0$, yields an agent's dimensionless energy liberation rate
\[
\dot{E} = \zeta (1-A) + \frac{1}{\alpha^2},
\]
where $\zeta = \frac{k_a g_0}{f_0 v_0}$.

Order-of-magnitude estimates show that $\beta$, $\gamma$, and $\zeta$ are $O(1)$, so we set them to unity. 
Again, more details on the derivation of this model can be found in Ref. \cite{mcgrath_microscale_2025}.
Each agent is assigned an inherent velocity-dependent unbinding parameter $\alpha > 0$, or muscle's concave force-velocity relation at the macro-scale, as the primary control parameter which, over generations and through mutations and selection, is evolved in the ABM.

\subsection{The agent-based model}

To study the evolution of actin--myosin nonlinearities $\alpha$ under competition and selection across generations, we implement an agent-based model (ABM) in which each agent actuates through a spatial environment using the minimal muscle model defined by Eqs.~\eqref{eq:A}--\eqref{eq:E}. 
Each agent possesses a phenotypic trait $\alpha$ that determines its velocity-dependent detachment dynamics and thus its actuation properties.

Simulations are performed on a two-dimensional periodic domain of size $L \times L$. 
At initialization, $n$ agents are placed at random positions drawn from a uniform distribution over the domain. 
Each agent begins with an internal energy reserve $E(0)$ and a trait value $\alpha$ drawn uniformly from $[1,10]$, reflecting the range of $\alpha$ most commonly observed in nature. 
Agents convert stored internal energy into motion according to the dynamical system in Eqs.~\eqref{eq:A}--\eqref{eq:E}. 
Their positions in the domain evolve according to $\dot{\mathbf{x}} = V\,\hat{\mathbf{r}}$, where $V$ is the normalized contraction velocity from Eq.~\eqref{eq:V} and $\hat{\mathbf{r}}$ is the unit vector pointing from the agent toward the nutrient resource.

At the start of each generation, a nutrient source is placed at a random position in the domain drawn from a uniform distribution. 
All agents orient toward this resource and consume internal energy according to Eq.~\eqref{eq:E} while actuating toward the nutrient by Eqs.~\eqref{eq:A}, ~\eqref{eq:V}. 
Agents whose energy reserves reach zero before arriving at the nutrient die and are removed from the population.

The first agent to reach the nutrient consumes it and fully replenishes its internal energy reserve. 
This agent produces an offspring whose trait mutates according to
\[
\alpha' = \alpha + \delta, \qquad \delta \sim \mathcal{N}(0,\delta^2),
\]
where $\delta$ controls the mutation strength. 
This offspring is placed at a random spatial location within the domain. 
Because agents tend to accumulate near the nutrient at the end of each generation, all agents are then displaced by a small random distance to prevent artificial spatial clustering and remove positional correlations between successive generations.

After nutrient consumption, reproduction, and removal of any dead agents, a new nutrient source is placed at a random location and the next generation begins. 
This cycle of competition, reproduction, and death is repeated for up to $10^4$ generations or until population extinction. 
Over time, selection acts on the trait $\alpha$, producing evolving population-level distributions $P(\alpha)$.

To characterize environmental resource availability, we introduce the distance scale $S$, defined as the maximum distance an $\alpha=1$ agent can travel before exhausting its total energy reserve $E(0)$. 
The ratio $S/L$ therefore determines the effective energetic constraints of the environment: small $S/L$ correspond to energy-limited environments in which agents cannot fully traverse the domain, whereas large $S/L$ correspond to abundant environments where nutrients are readily accessible.

For each parameter pair $(S/L,\delta)$, we perform at least $150$ independent simulations, each evolving up to $10^4$ generations unless extinction occurs earlier. 
Population-level phenotypic distributions $P(\alpha)$ are estimated from the resulting ensembles using kernel density estimation (KDE) to generate continuous, normalized probability distributions.
KDE is implemented using the \texttt{gaussian\_kde} function from the \texttt{scipy.stats} module in Python.

\subsection{Jensen–Shannon divergence calculations}

Throughout this study, we compute the Jensen--Shannon divergence, $\mathrm{JSD}$ --- a symmetrized and bounded variant of the Kullback--Leibler divergence --- as a measure of statistical distance between different phenotypic distributions of the trait $\alpha$. 
For two distributions $P$ and $Q$, the Jensen--Shannon divergence is defined as
\[
\mathrm{JSD}(P \| Q) = \frac{1}{2} D_{\mathrm{KL}}(P \| M) + \frac{1}{2} D_{\mathrm{KL}}(Q \| M),
\]
where $M = \frac{1}{2}(P + Q)$ is the mixed distribution and $D_{\mathrm{KL}}$ denotes the standard Kullback--Leibler divergence. 
The quantity $\mathrm{JSD}(P \| Q)$ is bounded between $0$ and $1$, with smaller values indicating greater similarity between distributions $P$ and $Q$.

In this study, we use $\mathrm{JSD}(P \| Q)$ for two purposes: (1) to quantify the similarity between steady-state distributions obtained in simulations $P_{\mathrm{st.st.}}$ and the natural distribution measured from biological tissue samples $P^*$, and (2) to quantify the evolutionary exploration of distributions between generations $i$ and $i+\tau$ as the agents adapt over time.

First, as shown in Figs.~\ref{fig:with_and_without_mutations}, ~\ref{fig:outcomes}, we compute $\mathrm{JSD}$ between the steady-state distribution realized in simulation $P_{\mathrm{st.st.}}(\alpha)$ and the natural distribution $P^*(\alpha)$ obtained from a collection of biological tissue samples centered around $\alpha^*$. 
Lower values of $\mathrm{JSD}(P_{\mathrm{st.st.}} \| P^*)$ indicate that the emergent steady-state distribution more closely matches the distribution observed in nature.

Second, as shown in Fig.~\ref{fig:mechanisms}, we use $\mathrm{JSD}$ to quantify the temporal dynamics of the evolving distribution across generations. 
Analogous to a mean-squared displacement calculation, for each simulated distribution of $\alpha$ at generation $i$, $P_i(\alpha)$, we compute its statistical similarity to the distribution at a later generation $i+\tau$, $P_{i+\tau}(\alpha)$, and average over all $i$.
This yields the time structure function $\langle \mathrm{JSD}(P_i \| P_{i+\tau}) \rangle_i$.
Repeating this procedure for all generation intervals $\tau$ quantifies how the population distribution evolves and adapts over time. 
As in Fig.~\ref{fig:mechanisms}, we plot $\langle \mathrm{JSD}(P_i \| P_{i+\tau}) \rangle_i$ versus $\tau$ on a log--log scale and interpret the resulting slopes analogously to mean-squared displacement in Brownian motion: a slope of $1$ indicates diffusive evolutionary exploration, a slope of $0$ suggests evolutionary caging, and intermediate slopes $0-1$ correspond to sub-diffusive, directed evolutionary dynamics.

\section{Acknowledgments}
We would like to thank Matt Visomirski and Ilya Beskin for helpful discussions. This work was supported by the National Science Foundation under Grant No. DMR-2144380.

\small{
\bibliographystyle{unsrt}
\bibliography{abm_bib}
}
\end{document}